# A Comprehensive Study on Pedestrians' Evacuation

Danial A. Muhammed (✉)
University of Sulaimani, Sulaymaniyah, Kurdistan, Iraq
danial.muhammed@univsul.edu.iq
Soran Saeed
Sulaimania Polytechnic University, Sulaymaniyah, Kurdistan, Iraq
Tarik A. Rashid
University of Kurdistan Hewler, Hewler, Kurdistan, Iraq

Cite as: Danial Muhammed, Soran Saeed, Tarik A. Rashid (2019). A Comprehensive Study on Pedestrians' Evacuation, International Journal of Recent Contributions from Engineering, Science & IT (iJES), Vol. 7, No. 4 : DOi: 10.3991/ijes.v7i4.11767

**Abstract**— Human beings face threats because of unexpected happenings, which can be avoided through an adequate crisis evacuation plan, which is vital to stop wound and demise as its negative results. Consequently, different typical evacuation pedestrians have been created. Moreover, through applied research, these models for various applications, reproductions, and conditions have been examined to present an operational model. Furthermore, new models have been developed to cooperate with system evacuation in residential places in case of unexpected events. This research has taken into account an inclusive and a 'systematic survey of pedestrian evacuation' to demonstrate models methods by focusing on the applications' features, techniques, implications, and after that gather them under various types, for example, classical models, hybridized models, and generic model. The current analysis assists scholars in this field of study to write their forthcoming papers about it, which can suggest a novel structure to recent typical intelligent reproduction with novel features.

## 1 Introduction

Some emergency cases will not be controlled easily and will be obstacles in front of the evacuation process because of perplexity, dread and even uncertainty and uneasiness to mass dwellers [1]. There are many factors that affect evacuation processes such as the surrounding, how they react with each other, and various environmental conditions, that is why many evacuation approaches occur like protective, preventive rescue and constructive evacuation [2]. The issue of evacuation proceeds to the crowd's movement and it is affected by the physical and social environment, such as the high degree of danger, pressure, and lack of data, which is a mixture of environmental hazards; population demographics and the attendee's conduct [3]. A crowd is gathering of a group of people [4] that has many features, during simulation a number of potential behaviors are anticipated [5], and simulation is a way of guessing behaviors through answering for the "what-if" conditions [6]. Hence, Crowd evacuation is a way to aping the behavior of participants in the same situation [4], in the last 20 years many types of research have been done, practicing evacuation have been considered to lower the damages; deaths and injuries in emergency situations involving pedestrians [7-13]. The sole purpose of the investigations is to improve a managing emergency situation that is why many models are enhanced to see how people react in different scenarios in emergencies [14-20]. One way of finding solutions is modeling thus after the careful examination because of the originality of the procedure it has been renamed a model [4]. The models could be divided into three groups; classical, hybridized and generic models, each of which is subdivided. These models have been used to explore crowd evacuation in normal and emergency situations.

This research has been conducted with the aim of First is to gather a huge number of papers about different 'applications' conducted on the evacuation of the pedestrians. The second is to examine characteristics, 'techniques' and the indications of various applications. Third, is to learn from the first two aforementioned points to decide to scheme a novel smart and dependable model to pretend attendees' 'appearing emergency' conducts and evacuation efficacy when an area is in need of emergency evacuation. Thus, the current research sheds light on the pedestrian evacuation specifically and generally. Yet it is a chance for the scholars to gain relevant information with ease and decide about how their forthcoming papers be directed.

The current paper will contribute to the literature by benefiting from the existing studies to assist scholars to utilize it in different cases. It can guide them to master their forthcoming expected studies, firstly. It can assist the researchers and those who specialize in design relying on the obtained results which were obtained through investigation of the past studies with the intention of making a decision in a better way for designing and implementing a novel smart 'simulation model' comprising recent capacities, secondly.

The following is the structure of this research: In the second section, 'evacuation models for the crowd' is presented. In the third section, the previous models and the relevant methods in various applications are demonstrated. The final section contains the conclusion and suggestions for further research.

## 2 Evacuation Models for Crowd

The crowd is performing a bunch of people together [4]. It is the only condition in which reverse could be the alternative of panic rather than being touched [21]. In 1895 presenting La Psychologie des foules by LeBon was the starting of researchers' concern for crowd dynamics [22]. On the other hand, Helbing in 1991 presented a widespread work, which was one of the attempts for such reason to display the motion of pedestrians [23]. Scientists, until 2001, had previously created relating models going for alleviating clog and obstacle wonders dependent on experimental information [24]. Meanwhile, different fields studied crowd dynamics when pedestrians' dynamic was presented [25]. Irregular movement, the effect of congestion and the occurrence of own-arrangement were specified [26]. Numbers of applications were simulated via this crowd evacuation model [25-30]. For displaying crowd evacuation from building various methods developed, such as cellular automata method, social force method, lattice gas method and agent-based method [31-33]. Hence, these methods based on the ability to know the detail of the individuals in the crowd covered via three different models macroscopic model, mesoscopic model and microscopic model [34, 35]. On the other hand, different hybridized methods were developed, such as zone based, layer based, and sequentially based [36]. Another method generic framework was presented from the previous mentioned hybridized models [34].

Figure 1 shows an overview of the developed models for crowd evacuation. The crowd models are categorized into three main models; these are a classical model, a hybridized model, and a generic model. Each model has its own approaches to investigate the flow of people and their behaviors during the evacuation process.

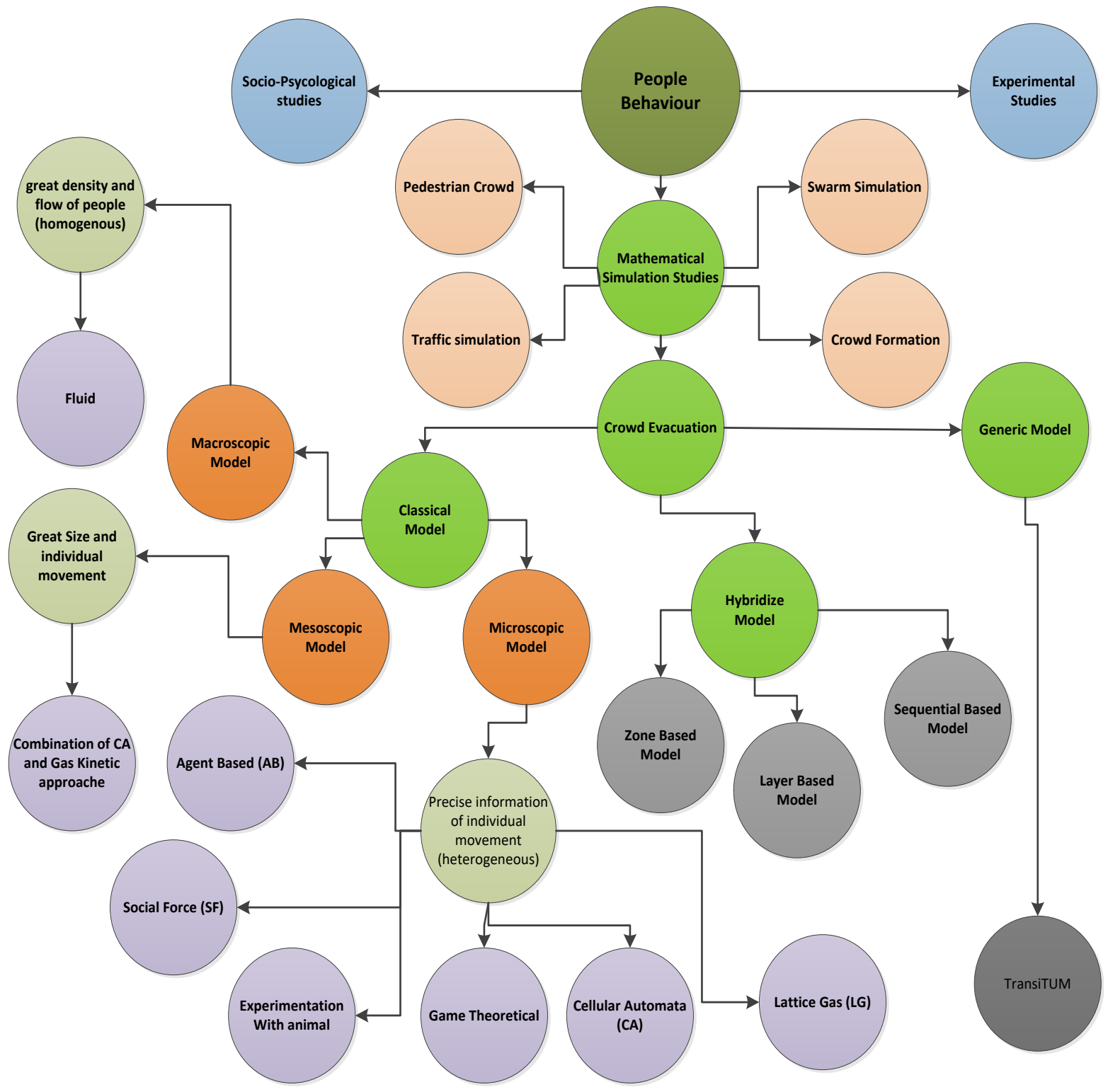

**Fig. 1.** An overview of the developed models for crowd evacuation

# 3 Models and Their Approaches in Different Applications

In this section, the models and approaches are described, and emphasis on highlighting the features, techniques, and implications of current simulation models. Table 1 shows information for future use.

## 3.1 Classical Models

The classical model can be divided into three different models; macroscopic, mesoscopic and microscopic. Each model was to design different approaches to know how humans move and behave during movements from one place to another of the specified area. These models are described in the following subsections:

**Macroscopic Model:** Macroscopic is one of the classical models and with such a model flow of people is noticed and individual features are neglected due to dealing with the homogenous people. Figure 2 illustrates the macroscopic model. In the macroscopic model, the fluid dynamic was designed.

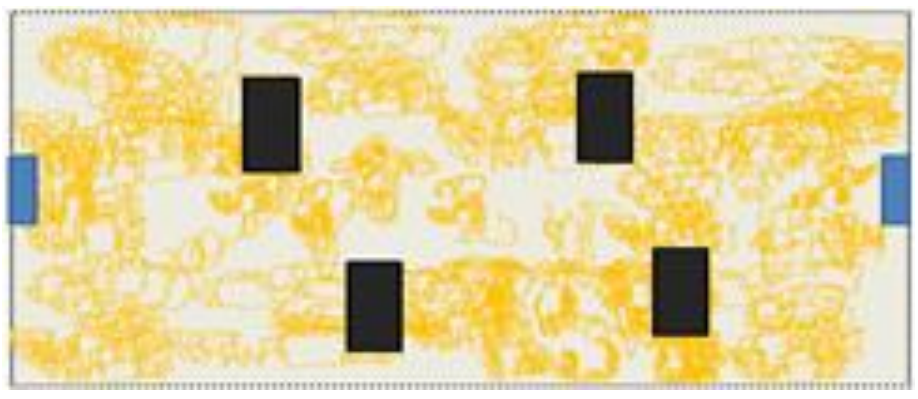

**Fig. 2.** Macroscopic model

In previous decades, fluid-like characteristics had been represented as a pedestrian crowd. There were numbers of connections between fluid and pedestrians, for instance, movement on all sides of the obstructions shows follow "streamlines", so, it was not an unexpected situation, especially, such as the premature models of pedestrians, which is vehicular dynamics took motivation hydrodynamics or gas-kinetic theory [37-40]. Henderson believed that a person on foot flocks acts comparably to gases or fluid [41]. Bradley estimated that the Navier– Stokes conditions administering smooth movement could be utilized to depict movement in groups at high densities [42]. Helbing et al. abridged that at medium and high densities, the movement of a person on foot flocks demonstrated some hitting analogies with the movement of fluid. For example, the impressions of people on foot in snow seem to be like streamlines of liquids or, once more, the surges of walkers through standing groups are practically equivalent to riverbeds [43]. Liquid powerful models portray how thickness and speed change after some time with the utilization of halfway differential conditions [44].

In 2002, Hughes designed a continuum theory for the flow of pedestrians. The present hypothesis is intended for the advancement of general strategies to comprehend the movement of vast groups. Nevertheless, it is additionally helpful as a prognostic instrument. The manner anticipated by these conditions of movement is com-pared with the existing reaction for the Jamarat Bridge close Mecca, Saudi Arabia [45]. In 2003, Hughes built up a continuum model distinct from a classical fluid in light of the property that a group has the ability to think, fascinating new physical thoughts are associated with its investigation. This property made many intriguing applications scientifically controllable. To do this, models were given in which the hypothesis had been utilized to give conceivable help with the yearly Muslim Hajj, to comprehend the Battle of Agincourt, and shockingly, to find obstructions that really increment the stream of people on foot over that when there are no hindrances present [46]. Moreover, in 2004, Colombo and Rosini displayed a continuum show for the person on foot stream to represent normal highlights of this sort of stream, namely, a few impacts of fear. Specifically, this model depicts the conceivable over compressions in a group and the fall in the surge through an entryway of a freezing swarm stick. They considered the circumstance where a gathering of individuals needs to leave a passageway through an entryway. On the off chance that the maximal surge permitted by the entryway is low, the progress to freezing in the group moving toward the entryway may almost certainly cause an emotional decrease in the real outflow, reduction the outflow much more [47].

**Microscopic Model:** There are several old models of which microscopic is one. Within which everything is realized accurately such as full information about individual and individual manners. Nevertheless, it is not perfect in examining the huge number of attendees. Figure 3 illustrates the microscopic model. Several objects are designed in microscopic such as cellular automata, lattice gas, social force, agent-based, game theory, and experimental approaches. Details about cellular automata and its applications are demonstrated below:

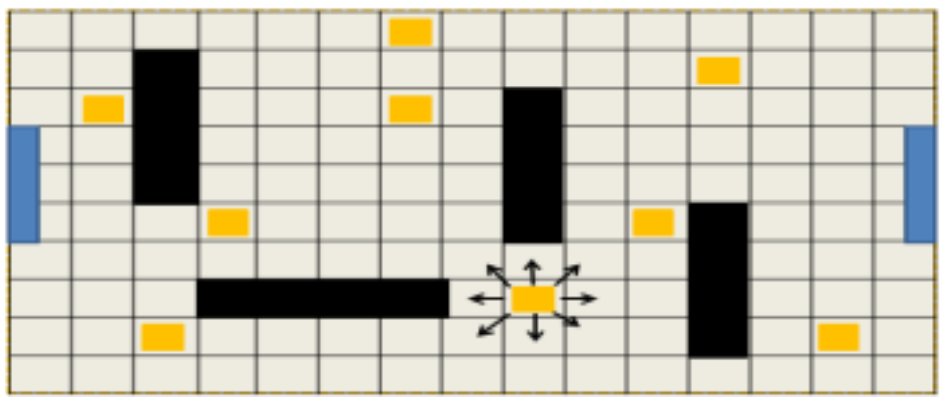

**Fig. 3.** Microscopic model

*Cellular Automata*

The accurate invention of physical methods in which 'time and space' are a remote and liable set of dissimilar values being approved inside the corporeal dimensions is called cellular automata. Cell automation includes a normal identical network, which is to some extent perpetual in grade with a different variable at every position (cell). The status of each cellular automation is mostly based on the approximations of the total reasons for all sites. The development of cellular automation in distinct stages, with the speculation of the 'variable' at a site undergoing influence of the reasons at endpoints 'in its neighborhood on the' start of the previous procedure. For the area of site (cell) it is essential to take into account two things: the 'site' and the 'neighboring locales'. The causes at all the sites are to be up to date together at the same time in order, in the light of the speculations of all the causes in their neighborhood at the start of the previous process, and for the distinctive preparation of local instructions in the corporeal abilities. They have been connected and reintroduced for a wide assortment of purposes and alluded to by an assortment of names, including tessellation automata homogeneous structures, cell structures, cellular structures, and iterative arrays. Von Neumann and Ulam were the ones who introduced cell automata first, which they called it cell spaces, like imaginable idealizing of 'organic' outlines (Von Neumann, 1963, 1966), which has a unique enthusiasm behind showing 'natural self-multiplication'. For a variety of reasons they have been linked and reinstated and referred to by a variety of names such as tessellation automata homogeneous structures, cell structures, cellular structures, and iterative arrays [48].

In the last two decades, cellular automata models have been created to consider an evacuating group of individuals under different circumstances. These models can be categorized into two groups. The first depends on the associations among situations and walkers. For example, in 2002, Perez et al. illustrated a cellular automata model to study pedestrian exit dynamics that distributed within a single room and content to leave through an experienced way out at the earliest possible time. The possible direction of pedestrians' movement was cardinal movement (or forward, backward, left, and right) which relied on the empty neighbor grid and specify coercion to their corporeal ability relation with neighbors and movement in conformity with ordinary instructions. This investigation presented the arching behavior due to the jamming effect at the way outs. Moreover, in the simulation of the way out output it observed various features, such as flowing and disorderly intervention. Furthermore, widths of the rooms' way out that create the possibility of pedestrians' exit at the same time caused to pedestrians leave the room in various sizes of bursts [49].

In 2002, Kirchner and Schadschneider demonstrated leaving process imitation utilized a new presented cellular automaton model for pedestrian dynamics. The idea of chemotaxis utilized in this model, which used a bionic method to define communication between pedestrians. In this research, some relatively simple conditions were examined, for instance, leaving a big room with a single or a couple of way outs. In addition, it was found that changing in dimensions of the model can define manner from regular to panic in various forms. Furthermore, it is discovered that for accomplishing best leaving times an appropriate amalgamation of herding manner and utilization of way out familiarity was essential [50]. In 2003, Kirchner et al. enhanced a new proposed cellular automaton model for pedestrian dynamics with adding friction parameters. This research investigated the effect of pedestrians' collisions. Friction parameter applied to prevent the possibility of moving conflicted participants into the same space at the same time step. Hence, this type of conflict is possible, but eliminating such situation is crucial in the precise definition of the dynamics. Besides, creating local compression among pedestrians in the model due to the friction parameter made the model have a role in an area with great density. Large room with a single exit door used for the evacuation simulations' experimentations. From the experimentation's result, it was discovered that the friction parameter in arching behavior participated in both of the quantitative influences and qualitative fluctuate [51].

In 2005, Yang et al. utilized a two-dimensional Cellular Automata model in mimicking leaving the process with kin manner. Within the real leaving process several attraction occurrences, such as confusion, congestion, assembly, step back and waiting, pretended due to the difference in constructing of the building, the organizing participants, choosing a path and the interesting for the kin manner. From the simulation results noticed some times walking in mass could be safe, there was no difference in leaving with one exit door and two exit doors from the aspect of kin manner, leaving efficiency greatly affected by sub-groups number and size of the sub-groups, as well as waiting and steps back decreased leaving efficiency [52]. In 2005, Li et al. presented a distinctive procedure based on human behavior to make the rules more practical for pedestrian movement, and then via assuming the bi-direction walker motion in a corridor bottom-up and top-down walkers' motion was illustrated, as well as probability of the influence of walkers' swapping locations was identified [53].

In 2006, Zhao et al. offered a two-dimensional cellular automata model to imitate participant leaving through exit dynamics. This study emphasized on two features, way out width and door partition. Hence some convenient aspects appeared, such as the width of the way out ought to be higher than a critical value, and door partition ought to be medium not too large and not too small. Moreover, One way out's width increment resulted in reducing the flow out for each unit width, nonetheless entire flow out greater than before. Whole way out's flow out was a cumulative nonlinear function of the way out width. Furthermore, way out width assessment did not effect on door partition's best value, and way out design had better be balanced. These aspects improve the efficiency of building design [54]. In 2006, Georgoudas et al. utilized a two-dimensional cellular automata model and applied a computational intelligent technique to examine pedestrians dynamic within a wide space. In addition, this recommended model differs from the previous ones when the model treated with a heterogeneous crowd. In fact, replying heterogeneous parts to the instruction, artificially arranged manner in the crowd, and made pedestrian attain one of the way outs. Finally, characteristics of pedestrians' actions utilized to examine different assumptions such as pedestrians' collision during the leaving process, collective effects, suspending issues, and fixed and movable obstructions [55].

In 2007, Varas et al. utilized a two-dimensional cellular automaton model to imitate the process of leaving from a single and double door classroom with complete ability. In this study, to each grid, the structure of the room, obstructions spreading, floor field were measured. Moreover, the effect of panic as a counted dimension was calculated which % 5 possibilities of not moving. The model applied the random selection to cope with collision issue. Therefore, the proposed model changed into non-deterministic via these characteristics. From the simulation result, it was clearly observed that the best locations of the door and evacuation efficiency were not enhanced by substituting double door with two distinct doors. Finally, for the evacuation time, a number of persons and way out width were considered due to suggesting simple scaling law [56].

In 2007, Yamamoto et al. demonstrated a real-coded cellular automata (RCA) model dependent on real-coded lattice gas to simulate the left from an area with one way out of different widths. The method of changing the pedestrian's position was exposed. In the former developed cellular automata model, the movement was partly simple, while they mimicked the over straight line movement and avoiding the perverse direction. Hence, observing the precise duration of evacuation was difficult. In this developed model, pedestrians allowed to move in the desired direction and measured the rational duration of the evacuation. From the simulation results, congestion observed at the exit of the big room. In fact, a critical number of pedestrian investigated who made congestion.

The correlation between the number of people in the room and the total evacuation time was achieved. Two regions region 1 and region 2 were tested. Inside region 1, however the number of pedestrians increased, the evacuation time remained steady. In contrast, inside region 2, people in the room needed more time to evacuate when an initial number of people increased [57].

In 2011, Alizadeh put forward a CA model to examine the procedure of evacuation in a place which was provided with 'obstructions' which had various configuration of the place, like places of exit and obstruction, 'the width of the exit, light' of the place, psychological status of the evacuee and the dispersal of the people gathered. Its influence was clearly seen in the process of evacuation. A restaurant and a classroom were taken as a case of this model. The way the evacuees distributed, 'location and with of the door on of the evacuation' discussed and production of the model was made ready for comparison with some motionless models [58]. In 2014, Guo, Ren-Yong made a model relied on 'CA with a better separation of the area and advanced speed of walking' to show going away of people who are walking from a place with one door for exit. Two factors affected the shape of the people gathered during the experiments; 'the advanced speed of walking and the separation of area' interval of people at different places and the efficacy of the people who left their houses shown through clocks. Moreover, the connection of 'width and flow of the exit' was demonstrated through this archetype [59].

In 2015, Li and Han proposed a model for simulating pedestrian evacuation relied on widened cellular automata to support various behavioral tendencies in people. Understanding and violence were two of the selected social tendencies to be looked at through this archetype. When examining this simulation, social constraints and pedestrians flow orders were confirmed. The results of the study show that evacuation time does not increase with an individual's knowledge and does not decrease when the individual's condition is noticed as aggressiveness. It is quite obvious that when the individual avoids aggressiveness in his conduct, the best type of evacuation will be recorded [60]. In 2018, Kontou et al. made a model of crowd evacuation on cellular automata (CA) parallel computing tool to simulate and evaluate manners and different features of pedestrians in the evacuation area; including disables. The simulation process was made in a school where disables existed. A center of education in Xanthi, which contained disable people, was selected for the simulation process. With observing and prevailing earthquake, the school organized security application; the total time of the emptying was noted. Lastly, suggested archetype through the experimental data validated and there was a suitability implication to the particular location [61].

*Lattice Gas Models*

In 1982 by Fredkin and Toffoli and in 1983 by Wolfram Lattice gases were promoted, which is a unique instance of cellular automata [62-64]. The individual on the grid of lattice gas models is measured as an active element. Possibility and measurement were considered to help these models to investigate individuals' crowd characteristics [44]. Individuals are fixed with L × W in this model, one individual is for one location. Based on executing a biased random walk with no back steps, the individuals move to a special direction, and available locations are allowed solely [65].

In 2001, Tajima et al., used lattice gas models of biased-random walkers to pretend walker channel stream at a bottleneck under the open boundaries. Then they noticed changing free flow into chocking flow under serious appearance density, filling flow proportion and changing the measure of density, and connection between flow rate and scaling law [66]. In 2002, Itoh and Nagatani presented a lattice gas model of pedestrians to simulate moving of the gathering of people between two halls through a door, and they noticed an ideal admission time for moving the viewers. The time has effect on the gate when visitors want to enter the hall because with decreasing the admission time to under optimal time jammed state occurs and the viewers have no ability to enter the inner hall while arrival pedestrians stopped by the departure pedestrians from entering [67]., in 2003 by Helbing et al. They made experiments and simulations for leaving process from a classroom. For the experimentation, they utilized video cameras for recording leaving students from a classroom and leaving time of each student recorded. Alternatively, for the simulation, they applied lattice gas model of pedestrian flows to compare with the experimentation results. They noticed disorganization specification empirically is well repeated in the evacuation process, and initial location has a major role in leaving time. Jamming or queuing state at the exit has a great effect on increasing leaving time [68].

In addition, this lattice gas model was utilized to think of group evacuation under various circumstances. For instance, in 2004, Nagai et al., made experimentations and simulations to present leaving the process in a room without visibility with a number of exits. In the experimentations, sightless students wearing eye covers imitate individuals in a room without visibility. Additionally, the video camera was applied to record the evacuation of confused students, and then the student path and leaving time were evaluated. On the other hand, students' detected manners were mimicked via the extended lattice gas model wherein sightless students are simulated through biased random walkers. Further, the mean value of the leaving time and students' leaving dynamic patterns were measured and made a comparison with the output of the experimentations. In this study, the exact emphasis was on the leaving time distribution [69].

In 2005, Nagai et al., made experimentations and simulations to show two types of counter-flow of students going on all fours in an open boundaries channel. In the experimentation, video camera utilized for recording and capacities of each student arrival times were calculated. Experimentally features of counter-flow were elucidated.

This research made a comparison between pedestrian counter-flow and the counter-flow of students on all fours. In this study, lattice gas simulation was applied to imitate the experiment, and a biased random walker utilized to pretend students crawling [70]. In 2006, Song et al., they made up a new lattice gas model ''multi-grid model'' due to presenting force concept of social force model into a lattice gas model. Therefore, better lattice participated which made the walkers reside in more than one grid, and constructed instructions for walkers and walkers and structures. This new model was used to simulate leaving walkers from a big room with an exit door, thus, the effect of collaboration force and drift factor on leaving time were evaluated. Finally, a common limitation of the two factors on the leaving process was discovered [71].

In 2007, Fukamachi and Nagatani Studied sidle influence on counter-flow pedestrian and investigated manners of sidle walkers within the crowd in the counter-flow pedestrian. Individuals within the crowd change his moving into sideways in order to be far from congestion and barrier. In this study, the influence of sidle investigated with the enhanced biased random walk model. Three models were demonstrated; 1) face to face usual pace, 2) only sidelong pace, 3) during crowd and barriers change usual walk into sideways, and get back to usual walk when mass left. They noticed the usual pace was slower compare to sidelong pace due to rising congestion, and emergent jamming state in the transference points. In the model number 3, jam cluster nearby middle of the channel extremely fluctuating near the jamming transference point [72], also, various ways joined with lattice gas models to deal with leaving process research.

In 2012, Guo et al. created a varied lattice gas model via utilizing both models of cellular automata (CA) and mobile lattice gas model (MLG model) to simulate evacuation processes during an emergency. Inside this model concept of local population density presented, and in this new model, this concept with a factor of exit crowded degree applied to update rule. Besides, drift D that is a significant parameter has an impact on the evacuation process and can be changed with considering the presented concept. The nonlinear function of the corresponding distance used to define communications, such as friction, attraction, and repulsion between every two walkers and walkers with building dividers. When spaces between them get smaller, repulsion forces increase severely. Simple characteristics of pedestrian evacuation, such as clogging and arching phenomena could be taken from numerical examples [73].

In 2013, Guo et al. offered an agent-based and fire and pedestrian interaction (FPI) model to investigate the leaving process during an existing emergency. It was thought that the environmental temperature field creates an effect on probability direction of the movement. Besides, the multi-grid method was applied to define decreasing speed by low transparency in the fire and pedestrian interaction (FPI). Hence, the authors created an extended heterogeneous lattice gas (E-HLG), model. Inside this new model factor of altitude was added to define the height location of lattice locations. Due to the model and experimentations, characteristics of the left in a terrace classroom were studied. Outputs from the extended HLG model were close to the experiments. In addition, leaving controlled due to the altitude factor, and the different decision of choosing evacuation paths and annoying high-temperature field causes to local jamming and clogging [74].

In 2016, song et al., created an evacuation scene based on cellular automata and a lattice gas model to simulate behaviors of selfless and selfish for the pedestrians during the evacuation and competitiveness behaviors, meanwhile to present the influence of them on pedestrians' strategies. Furthermore, some experimentation performed on the width of the building exit door and analyzed. Outputs of the simulation tests demonstrated that individuals with self-behavior caused more deficiency and rise evacuation duration. Conversely, sympathy caused to decrease evacuation duration and more collaborators. Finally, an important factor for the duration of the evacuation was the exit door width. When the size was less than six cells of the size of 50 x 50, evacuation time increased, conversely, the time was seriously decreased with increasing the width. However, this would be no noticeable when the door exit width much more increased [75].

*Social Force Model*

In 1995, Helbing and Molnar presented that pedestrian movements can comply with 'social forces'. The movement of the pedestrian is controlled by the accompanying principle impacts, which are first, pedestrian needs to achieve a specific goal. Secondly, pedestrian keeps a specific separation from different people on foot. The third one is that pedestrian additionally keeps a specific separation from the edge of obstructions, for example, dividers. Fourthly, a pedestrian is some of the time is pulled in by different people or objects [76].

In 2000, to simulate fear conditions Helbing et al. built an alternative social force model. In this model combination of physical forces and socio-psychological with people mass behaviors referred [77]. In 2002, Zheng et al made a combination of the social force model (Helbing et al., 1995) and neural network as a model to simulate different conditions of walkers for collective behaviors [78]. In 2005, Parisi and Dorso applied the social force model, which presented by Helbing and assistants in 2000 to permits investigation of various levels of fear in evacuation within a single exit door room. In this research, presenting concept 'faster is slower' participated in changing manners and the rising chance of congestion suspensions, and also this concept with the blocking clusters made a robust connection, whereas, size effect for exit door was concisely debated [79].

In 2006, Seyfried et al. applied the adapted social force model which was presented by Helbing and assistants in 1995 to specifically examine the effect of various methodologies for communication between walkers, which are self-driven objects moving in a continuous space on the velocity-density relation's output. Consequently, they noticed it is possible to simulate the usual arrangement of the fundamental diagram when individual space and current speed are increased. Additionally, they present distant force has an effect on velocity-density relation [80]. Besides, the social force models are joined with different models to examine swarm departure. In 2006, Lin et al., suggested a system for evacuation crowds during emergencies via dynamic model to propose a standard framework for future studies. Carrying out standard func-tions was the main task, and emphasized on framework constancy and capable of being extended to propose new required tasks in the future. This study tried to enhance framework independency and better system execution. Hence, experimentations were executed in a certain construction to evaluate the effectiveness of the crowd evacuation, thus, in the experimentations results specifically interpreted crowd manner, construction, and density of the people were mean of the effect [81]. Later in 2007, utilized the social force model to make pedestrians be dynamic, and then examined 200 pedestrians leaving process from a room during a panic situation. Parameter such as υd, which was denoted aspiring speed for pedestrians to move, was applied to control panic levels. In this study, the effect of "faster is slower" concept according to the attempts with applying various forces was known. When υd efficiency of the evacuation starting to reduction swiftly and flaw rate reaches to the peak, exponential mass distribution changes into ''U-shaped'' [82].

In 2008, Guo and Huang proposed a mobile lattice gas model based on utilizing both social force model and lattice gas model benefits. The model specifies each pair walkers' communication and structure partition with pedestrian communication due to space and pace size in motion. The output of this emergency simulation model demonstrated 1) walkers' evacuation simple features, such as arching and clogging behavior and practically produced a mean of the evacuation time 2) load computation not as much as social force model and gain duration of the evacuation more precisely [83].

In 2011, Okaya and Takahashi utilized a BDI model to simulate the behavior of communications in the crowd, which usually occurred during evacuation. Inside such a model, evacuation behaviors were influenced due to people interactions. Hence, Helbing's social force model adapted in order to consider the intentions of the pedestrians. The output of the experiment's simulation demonstrated that as usual interactions among pedestrians due to congestion made the evacuation take a longer time. In addition, evacuation of family members together increased evacuation duration. Additionally, evacuation behaviors were influenced by directing the evacuation process [84]. In 2014, Hou et al. applied a modified social force model to simulate the influence of the number and location of the evacuation guiders on evacuation dynamics in partial visibility rooms. Inside this model, guiders who are qualified can identify the exit location precisely, and others compliance with the guiders' locations and instructions. Experimentations' output reveals for one exit, one or two guiders put a significant impact. Alternatively, for more than one exit position without adequate benefit from the whole exit, the evacuation gets slower. Consequently, it was obvious to increase the effect of guider on making evacuation faster, a number of exits with the number of evacuation guiders should be equal and guiders properly inside the room centralized of the multi-exits [85].

In 2017, Han and Liu applied a modified social force model involving an information transmission mechanism to simulate behaviors of walkers, when the majority walkers were unfamiliar with the evacuation location during a disaster. This improved model considers the approach of preventing collision and disappearing information. The difference between this adapted model and the previous model was this altered model defines the way of finding and selecting the correct direction, and the previous model was applied to eliminate the pedestrians collide. The output of the simulation demonstrated that due to information transmission mechanism walkers could determine the right motion direction, although walkers' real behavior could be simulated when emergency exists. Furthermore, there were different outcomes from the simulation was obtained to enhance the evacuation. Firstly, utilizing all exit door via the occupied extensively reduce time and rise efficiency of the evacuation. Secondly, using exits with more width completely causes the decreasing time of the evacuation and enhancing evacuation efficiency. Thirdly, in the start of evacuation walkers were restricted to select exits with greater width with fewer densities for their evacuation route. Lastly, at the start of the evacuation process essential directing was vital [86].

*Agent-Based Model*

ABMs are computational models that assemble social structures from the " bottom-up ", by reproducing people with virtual agents and making promise associations out of the task of principles that run connections among operators [87]. Bonabeau maintained the perspective of the following manner. In describing agent, the manner of mutual fear is an occurrence, which is growing due to the generally complex individual-level manner and cooperation among agents. Therefore, the agent-based model (ABM) appeared to be perfectly suited to give significant prudence into the method and prerequisites for fear and overcrowd by incoordination [88]. Nearly a a couple of decades, the ABM method has been utilized to contemplate crowd evacuation in different circumstances. ABMs compare to other methods, such as cellular automata, social force, lattice gas or fluid-dynamic models are commonly more computationally costly. Besides, dealing with heterogeneous people is considerably easier due to ABMs' capacity to enable every agent to have distinctive manners [44].

In 2004, Zarboutis and Marmaras demonstrated a method of modeling and simulating a metro system with an existing fire state in a tunnel. The ability of the simulation method to search for an effective strategy in the rescue was debated. This system included various subsystems and made multifaceted adjustable system. Hence, they established an agent-based simulation to support fitting dynamic illustration of the difficulties in the designed area, and 1) made the serious reliance and robustness of the system be captured by the designer, 2) described the intended rescue plan via the determined characteristics by the designer, 3) evaluated their proficiency. From the experimentations, output demonstrated different arrangements with different situations for metro personnel's activities [89]. In 2005, Braun et al., dependent on social force model exhibited an agent-based model (Helbing et al., 2000) to simulate effects of various floors, walls, and obstructions on agents, and cooperation among agents in emergency conditions. In the model, XML script was utilized to define the probability of reproducing various scenarios. Danger incident exhibited and visualized, multifaceted environment measured alarms organization were spread and considered through the environment, which led to reducing dead agents' numbers [90].

In 2006, Toyama et al., dependent on cellular automata suggested an agent-based model (ABM) denote various walkers' features, for example, room geometry knowledge, speed; gender, obstacle avoidance behavior, and herding behavior. They investigated the effect of various designs, various number of pedestrians in groups, and features on pedestrians dynamics and system's macroscopic manners [91]. In 2007, Pelechano et al. introduced High-Density Autonomous Crowds (HiDAC) model, which was a multi-agent model. This model relies upon psychological and geometrical instructions while it is a parameterized social force model. It very well may be adjusted to mimic different kinds of the crowd, extending from the high-density crowd under quiet conditions (exit from a cinema after a movie) to extreme fear circumstances (fire leaving) [92].

In 2012, Simo et al. demonstrated a model, which employed the social force model extensively, and agents' movement defined via Newtonian manner to simulate counter-flow conditions for mediators' behavior, which attempt to eliminate the colliding with approaching mediators. Inside this model, mediators noticed the moving path of head mediators, and their activities were selected based on that observation [93]. In 2012, Ha and Lykotrafitis applied self-moving particles system and movement was directed via social force model to simulate how various conditions, such as, multifaceted structure of the building, size of the room exit, main exit size, friction factor, and preferred speed influence on the evacuation time in which conditions enhancement occurred in evacuation efficiency. The output of the simulation demonstrated evacuation in a single room with small size for the exit, defined high speed takes longer time due to occurring overcrowding. With increasing, door size overcrowding disappeared. Friction had a significant effect on overcrowding, with minimizing the factor of friction overcrowding quickly disappeared.

On the other hand, for a floor with two rooms, one main exit door, and a hallway, a smaller the door room size may enhance evacuation efficiency. Because the number of the evacuee agents would be smaller from the room into the hallways, thus, the number of evacuees for the main exit would be smaller and agents can go out without occurring serious congestion. Conversely, with increasing room door size and agents' speed of the evacuation time decreased, while serious congestion occurred near the main exit door [94]. In 2018, Poulos et al. employed an agent-based evacuation model to simulate the school's staff and nearly 1500 children of an inclusive evacuation process executed for the whole city. This study emphasizes on kindergarten to 12th-grade school and examines the movements of various mediators. This simulation certified via comparing the real event, which shot video of the event, and expected a result from the developed model simulation, errors between the real and expected was the only % 7.6. Hence, output said that utilizing a mathematical model in evacuation for adapting logistical issues in an emergency arrangement is fair [95].

*Game Theoretic Models*

On the off chance that the intelligent choice procedure of the evacuees is reasonable, a game theoretic methodology can be embraced to display the choice circumstance [96]. In a game, the evacuees survey the majority of the accessible choices and select the elective that augments their utility. Every evacuee's last utility adjustments will rely upon the activities picked by all evacuees. Game is the determination of a cooperative state via a group of individuals, conceivable approaches of every individual, and the group of all conceivable utility adjustments. For one leave, the competitive behavior of the walkers in crisis departure could be deciphered in a game hypo-thetical manner [97]. For a few ways out, Lo et al., built up a non-agreeable game hypothesis display for the dynamic leave choice procedure of evacuees. The model inspects how the reasonable communicating conduct of the evacuees will influence the clearing designs. For the leave determination process, a blended procedure is considered as the likelihood of leave decision. The blended methodology Nash Equilibrium for the amusement depicts the balance for the evacuees and the blockage conditions of ways out [96].

In 2006, Lo et al., presented an original method to show the dynamic process of participants' way out choice. In a space with the extensive crowd, density participants manage his/her plan of evacuation on crowd's action, movement space to the way out, and influence of environmental motivation and way out familiarity. This plan is due to observing activities of participants and environmental conditions via other participants and reply to their familiarity to choose their egress path. This research presented a non-cooperative game theory framework. The model looked at how evacuation patterns and leaving time of an area with numbers of the way out influenced via the participants' collaborating manner. A merged procedure is considered for a way out determination as to the way out the choice possibility. The merged procedure Nash Equilibrium for the game defines the stability for the participants and the overcrowding conditions of way outs [96].

*Approaches Based on Experiments with Animals*

The utilization of creatures is another methodology for examining swarm departure. Tests in real departure freeze are troublesome, particularly with people in view of conceivable moral and even legitimate concerns. The elements of evacuation restriction are not totally comprehended in light of the fact that reviews have been largely kept to numerical recreations [98].

In 2003, Saloma et al., examined the dynamics of outflow fear in mice getting away from a water pool to a dry stage through an exit entryway. The investigation demonstrated how the engineering of the space in which they are kept affected in the manner of fearing groups. The investigation output exposed that for a basic inspecting interim their outflow manners concurred with the numerically anticipated exponential and power-law frequency distributions of the leave burst measure notwithstanding for brief time spans [98]. In 2005, Altshuler et al. used ants as a model of pedestrians to demonstrate herding behavior. In the experimentations, the ants were applied in one cell with equally placed two exits. Ants nearly used both exits to abandon in the same way when the situation was normal, however, with setting panic due to expeller liquid select one of the exits more severely. Hence, they modified the former hypothetical model, which comprised herding related to the panic element as the main element to simulate noticed dynamics outflow in detail. Moreover, from the experimentation outputs, emerged hypothetical models, preferred that during the leaving the process with the existing panic situation there are some common characteristics of the social manners between humans and ants [99].

**Mesoscopic Model:** Mesoscopic is one of the classical models and in these model movements of large size of people are investigated and somehow individual features are specified, Figure 4 illustrates the mesoscopic model. In mesoscopic, cellular automata and gas Kinetic approaches are combined. The following describes cellular automata and gas Kinetic approach:

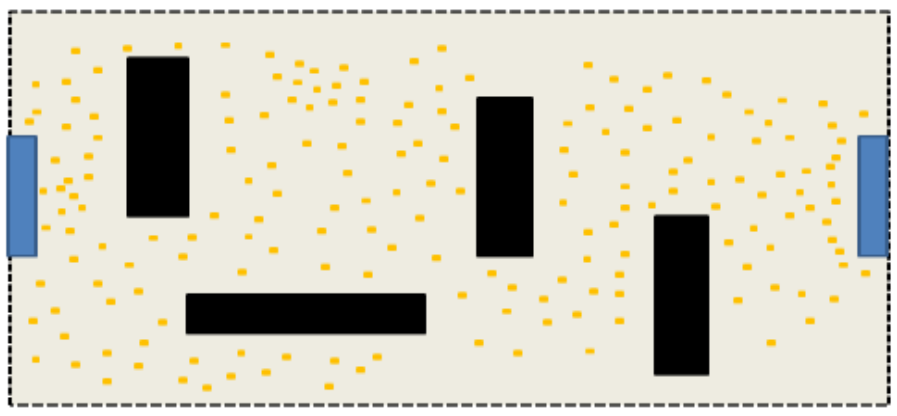

**Fig. 4.** Mesoscopic Model

*Cellular Automata and Gas Kinetic*

Cellular Automata (CA) with Gas Kinetic methodology made a mesoscopic model, which utilized into the motion of individuals observation. Besides, this model presents and imitates the great size of the group [100]. CA is a model, which is divided into numbers of grids; every grid has adjacent and different state [4]. In addition, CA to interact with simulating the departure of agents depends on separation, distribution, and utilizes an irregular way. CA thinks about the collecting manner of the agents. A key part of the CA display is more suitable to speak to pedestrian stream in perspective of its straightness, flexibility, and effectiveness [101].

### 3.2 Hybridized Models

Via using both macro and micro models of the classical models, a model, such as hybridized models designed and it can be divided into three different models; zone based, layer based, and sequential based. These models deal with the area of the evacuation and motion of the participants during the evacuation. These models are described in the following subsections:

**Zone-Based Model:** In this methodology, the area of simulation is partitioned into numerous zones. In light of use needs, each zone is reproduced either for the microscopic or macroscopic model. Zone imitation under macroscopic procedure gives in the general stream of the group, though zone mimicked with microscopic model offers singular dimension practices perception. Largely, the proposed procedures run the two models all the while on pre-defined zones [102-104].

In 2011, Wei et al. utilized Hybrid Grid Simulation Infrastructure to simulate a leaving process of a high-density mass in an urban area. Three heterogeneous models had been built that were a computational microscopic crowd model, a pedestrian agent model, and a vehicle agent model to depict different parts and features of the big and compound simulated situation. From the output exhibited that suggested infrastructure is a feasible and capable method for big and compound simulation system [103]. In 2011, Sewall et al., for shared simulation of largescale vehicle traffic for virtual universes and enlarged airborne maps exhibited a different strategy, which dynamically combines continuum and discrete approaches. Accepting these two unique approaches at the same time in various areas takes into account an adaptable simulation system where the client can without much of a stretch and naturally exchange quality and effectiveness at runtime. They had employed this method to the mimic of extensive systems of vehicle traffic dependent on artificial engineered urban situations, real-world information, and accomplished more prominent than continuous execution [105].

In 2012, Anh et al. showed a hybrid modeling method for evacuation simulation to increase the speed of pedestrians' movement and worked on the arrangement problem of both micro and macro models. Initial outputs demonstrated that to simulate leaving strategy in road network via the hybrid model more effective than via micro model alone [102]. In 2012, Xiong et al. wanted to use the benefits of macroscopic and microscopic models together. These two models in a simulation worked concurrently and performed within various shared special partitions. Through each step of simulation execution, each model had self-governing. Nevertheless, for crowd entering to opposite partition was allowed for each of the models. Models could join at edge border via using the various interactions of aggregation and disaggregation to swap information. From the output, it was exhibited that this hybrid model was more efficient than the microscopic model, and made quality enhancement compare to the macroscopic model [104].

**Layer Based Model:** Accepting way of applying both micro and macro methods partly into various layers is another method to deal with mass imitation. These applied methods are used in the whole area of the imitation in order to determine plane mass movement and additionally motion forms of the agents in the mass. This new method for both distinct layers does the arrangement of the global path, evasion of local obstacle and other wanted manners of the mass [106-108]. Inside this proposed method, the macro method applied to mimic mass motion in accordance set of rules in the first layer and the mass motion manner from this layer goes to the second layer as input. Hereafter, in the second layer micro method is using to mimic motion individuals independently and with rising density protect the cost-effectiveness.

In 2008, Banerjee et al. displayed an augmentation of the layered knowledge strategy that is prevalent in the amusement business for adaptable group recreation. They noticed a few navigation behaviors could be applied effectively in this system. The central preferred standpoint of this system is broadened capacity, where new behaviors can be included by including separate layers, without influencing the current layers. The edge rates have been observationally demonstrated to be adequate to handle huge groups continuously. A few angles have been distinguished where this simulation framework can be moved forward [107]. A natural way exhibited to deal with direct recreation of virtual groups using objective coordinated route capacities. The methodology effectively showed for a wide assortment of reproduction to produce diverse plainly visible practices and characteristic looking movement designs, resolve clog and perform an objective coordinated route. Mostly, this methodology offers a straightforward, yet incredible technique to direct or control swarm reenactments.

In 2011, Patil et al. exhibited a natural way and employed navigation functions to deal with the direct simulation of virtual groups. This approach effectively showed to produce diverse macroscopic behaviors and natural looking movement designs, resolve to clog and perform goal-directed navigation. This procedure was an incredible technique for controlling and directing the crowd simulation process [108]. In 2012, Tissera et al. exhibited a hybrid simulation model to check behavior patterns in an emergency leaving. Both environmental (EsM) and pedestrian (PsM) sub-models shared inside the hybrid model. Constructing a synthetic location occupied with independent cooperative agents due to the combination of the model with the computational procedure. Authors made sequence investigations; for instance, check the environment to the individuals leaving that were behaviors was available for the ”adjacent door”. After that, check the effect of familiarity of the individuals into the environment, outside motivation to instruct the individuals was utilized to the other conceivable outflow exit. The behavior of people reacting to this improvement is expected to "get out the entryway quicker” [106].

**Sequential Based Model:** Like layer based hybrid models, another methodology is a sequential hybrid procedure, which additionally runs both large scale (macro) and small-scale (micro) models for the entire group. It first runs a large-scale model to direct the motion forms of group and after that applies a small-scale model to the same group for watching the individual manners. It executes the two models in a successive way where a synchronization technique is required to exchange the group state between the two modes [109,110].

In 2011, Park et al. demonstrated a hybrid framework for crowd simulation due to applying both continuum-based and agent-based methodologies. The model catches the dynamics of the crowd in a vast group besides the individual behaviors of every agent. From outputs of the performance demonstrated that their methodology made a good balance in big and great determination simulated environment. Another promising benefit of this methodology is map field development due to the capability of the model to link to any possibilities for the map field, which gives the ability to enlarge the continuum-based simulation with non-lattice based paths [36]. In 2013, Xiong et al. suggested a hybrid model due to utilizing both macroscopic and microscopic models to simulate crowd in dynamic environments. Movement tendency for the crowd was simulated via the macroscopic model. On the other hand, determining the velocity and moving direction was due to the microscopic model. According to the outputs of the simulation appeared there is a good performance to show the features of crowd movement and human [109].

## 3.3 Generic Model

Due to crowd density, the Simulation of an application nearby requires using most suitable software, the needed dimension of individual manners (corporeal, mental and collective), and execution time. Simulation software projects are reliant on fundamental models that cannot be changed according to end client necessity. Hence, the generic model would be an important need to give the ability to choose models on user selection for detecting different crowd dynamics [111]. The following describes the transit approach:

**TransiTUM Model:** The latest attempt exhibited to build up a conventional structure for multiscale coupling of walker imitation models for transition zones [112]. Grouping different models, such as mesoscopic and microscopic models need the autonomous of these models. Besides that, essential parameters, such as speed, current location, subsequent goal, max speed and so on could be moved between them via assisting a data file. The displayed model concentrated on autonomous of related models and in this manner can be connected to any mix of mesoscopic and microscopic models. With the assistance of an outer information record, models can openly exchange essential parameters (speed, current location, subsequent goal, max speed and so on) between themselves. It has employed the idea of transit area and relaxation zones to flawlessly move the people from one model to another. Therefore, walkers can enter from any points. Nonetheless, this starter progress in the direction of conventional coupling and multi-point entry to transition zone needs further examination.

**Table 1.** Highlighting previous models and approaches' features, techniques and implications of current simulation models (where **MT**: Models Type, **MTH**: Methods, **MD**: Models, **PNT**: Participants, **SST**: Simulation State, **IOB**: Investigated Occurrences and Behaviours)

| MT | | REF | MTH | MD | PNT | SST | IOB |
|---|---|---|---|---|---|---|---|
| Classical Models | Lattice Gas Model Applications | [66] | LGM | Microscopic | homogenous | Normal | free flow to the choking flow, bottleneck width caused to saturated flow rate and transition density scale |
| | | [70] | LGM | | | Emer-gency | Counterflow of people crawling on all fours, speed, jamming transition, and pattern formation |
| | | [71] | LGM | | | Nor-mal/Em ergency | Sidle effect, counterflow jamming transition and pattern formation |
| | | [67] | LGM | | | Normal | Shifting of the audience between two halls, and Jamming transition |
| | | [68] | LGM | | | emergency | Jamming (queuing), and effect on increasing escape time |
| | | [69] | LGM | | | | Communication of escape times and the exit configuration, explain blind people feature proper-ties |
| | | [72] | LGM and SFM | | | | Determine simple features of the social force model, such as, clogging and arching, |
| | | [73] | MLGM and CAM | | heterogeneous | | Impact of concept of local population density on drift D within the evacuation, friction, attraction, and repulsion |
| | | [74] | E-HLGM | | | | Altitude factor added caused to control evacuation, choosing evacuation paths, annoying high tempera- |
| | | | | | | | ture field caused to local jamming and clogging |
| | | [75] | LGM and CAM | | | | Selfless and selfish for the pedestrians during evacuation |
| | Cellular Automata Model Application | [52] | CAM and SFM | Microscopic | Homogenous | Emergency | Impact of kin behavior on enhancing evacuation efficiency |
| | | [54] | CAM | | | | Effect of exit dynamics on flux, arching |
| | | [49] | CAM | | | Emer-gency | Arching, streaming, disruptive |
| | | [56] | CAM | | | | Effect of obstacles |
| | | [51] | CAM | | | | Clogging, friction effects, arching effects |
| | | [53] | CAM | | | Normal | Bi-direction movement, moving up and down, self-organization |
| | | [55] | CAM | | | Emergency | Clogging and mass behavior, arching, fixed and moveable obstacles |
| | | [50] | CAM | | | | Herding behavior |
| | | [57] | CAM and LGM | | | Normal | Clogging, movement pedestrians (oblique direction to the grid) |
| | | [58] | CAM | | | emergency | Impact of distribution of the evacuees, location, and width of the door on time of the evacuation argued |
| | | [59] | CAM | | | normal | the shape of the crowd, duration of the individuals at various positions, efficiency of the evacuees expressed via two time indicators, the association between width and flow of the exit |
| | | [60] | CAM | | heterogeneous | normal | Familiarity and aggressive, evacuation time |
| | | [61] | CAM | | heterogeneous | emergency | disable children, evacuation time |
| | Agent Based Model Applications | [89] | ABM | Microscopic | Homogenous | Emer-gency | Rule-based behavior |
| | | [90] | ABM and SFM | | | Emer-gency | Behaviours (risk, decision, escape) |
| | | [92] | ABM and SFM | | | Normal/ Emer-gency | Respectful behaviors when desired (Queuing, Decision), agent interaction (pushing behavior, panic propagation, impatience, real time reactions to changes in the environ-ment) |
| | | [91] | ABM and CAM | | | Nor-mal/Em ergency | Obstacle avoidance and herding behaviors, pedestrian group different features led to different escape probabilities |
| | | [93] | ABM and SFM | | | normal | Counter flow conditions, observing the moving path of head mediators and selecting activities based on that observation |
| | | [94] | ABM and | | | emer-gency | Effect of different features on occurring overcrowding and evacuation time |

| | | | | | | |
|---|---|---|---|---|---|---|
| | | SFM | | | | |
| | [95] | ABM | | | emer-gency | Movements of various mediators. Supporting that mathematical model in evacuation for adapting logistical issues in the emergency arrangement is fair |
| **Social Force Model Applications** | [77] | SFM | Microscopic | Homogenous | Emer-gency | Provide insight for panic and jamming, Clogging, faster-is-slower, mass behavior, pressure, beginning of panics by counter flows and impatience |
| | [78] | SFM | | | Normal/ Emer-gency | Crowd impatience, the existing optimal ratio of impatient persons to patient persons of pedestrians made the move-ment be quick and easy |
| | [80] | SFM | | | Normal | Impact of required space and remote action on the funda-mental diagram |
| | [79] | SFM | | | Emer-gency | Faster-is-slower, clogging, the impact of the exit door size |
| | [82] | SFM | | | Emer-gency | Faster-is-slower, cluster mass distribution |
| | [81] | SFM and ABM | | | Emer-gency | Separating data into a block, mass behavior, crowd's evac-uation efficiency |
| | [83] | SFM and LGM | | | Emer-gency | Simple features of pedestrian evacuation (arching, clogging), average evacuation time |
| | [84] | SFM | | heterogeneous | emergency | communications in the crowd during the evacuation |
| | [85] | SFM | | Homogenous | | Effect of Evacuation leaders on the evacuation process |
| | [86] | SFM with the infor-mation trans-mis-sion mech-anism | | heterogeneous | | Unfamiliar pedestrian behaviors with the evacuation location |
| **Fluid Dynamic Model Applications** | | | | | | |
| | [45] | FDM | Micro-scopic | Homog-enous | Normal | The motion of large crowds |
| | [46] | FDM | | | Normal | crowd in motion, ''Thinking fluids'' behavior |
| | [47] | FDM | | | Emer-gency | Impact of over compression |
| Theory Model for game Applications | | | | | | |
| | [96] | Theory Model for game | Microscopic | Homogenous | Eme rgen cy | exit choice, the impact of evacuees' interaction on evacuation pattern and clearance time of a multi-exit zone |
| Animal Experimentations Model Applications | | | | | | |
| | [98] | Mice investigation | Microscopic | Homogenous | Emer gency | Self-organized queuing, diffusive flow, scale-free behavior in escape panic |
| | [99] | Ant investigation | | | Emer gency | Panic condition, herding behavior |
| **Hybridize Models** | **Zone-based Model Applications** | | | | | |
| | [105] | Zone-based | microscopic and macroscopic | heterogeneous | Normal | Leaving strategy in the road network |
| | [103] | | | | Normal | Leaving process of a high density mass urban area |
| | [102] | | | | emergency | increase speed of pedestrians' movement |
| | [104] | | | | Normal | Improve efficiency and enhancing quality |
| | **Layer-based Model Applications** | | | | | |
| | [107] | Layer-based | microscopic and macroscopic | heterogeneous | emergency | navigation behaviors, new behaviours can be included by including separate layers, handle huge groups continuously |
| | [108] | | | | | clogging and perform goal-directed navigation |
| | [106] | | | | | behaviour patterns in an emergency evacuation, environment, familiarity, and external motivation impacts |
| | **Sequential-based Model Applications** | | | | | |
| | [110] | Sequential-based | microscopic and macroscopic | heterogeneous | emergency | group behaviors and complex behaviors |
| | [109] | | | | Normal | crowd tendency, determine velocity, moving direction |
| **Generic model** | **TransiTUM Model Applications** | | | | | |
| | [111] | Transi-TUM | microscopic and mesoscopic with having external data file | homogeneous | Normal | Solving pedestrian transition between models, they can transit in different point. |

# 4 Discussion and Conclusion

Although the developed models relied on pedestrian evacuation methods have been used widely and completed, it is yet necessary to create models that precisely 'simulate people evacuation time' and existing manners in the time of emergency evacuation. It becomes necessary to review the past studies and make a distinction between dissimilar causes that influenced the manners of participants when emergency cases happen and similarly on participants' evacuation time.

This paper is a comprehensive and systematic survey of pedestrian evacuation which sheds light on 'applications' characteristics, methods, and implications of models methods. There are different categories for these models' methods and generic models. Forthcoming studies may benefit from this sorting of models and shedding light on their dissimilar features in several conditions and as a sort of guide to direct future studies. Finally, researchers can get benefit from results so as to be able to make better decisions in the system of evacuation and it will be a great support to enhance novel simulation models with novel abilities.

Relying on the literature review, the forthcoming studies will use a developed recent smart model. Moreover, the way fire is designed and implemented will be looked at. Other characteristics like the way fire spread out through the residential places as related to its location will be part of this, too. Additionally, the influence that the fire leaves on the agent's conducts will be explicated.

To finalize, the consequence of the fire and smoke will be written in the suggested model numbers of the demise, wounded or suffocated people.

## 5 Acknowledgment

The study is fully funded by the University of Sulaimani (UOS). The authors would like to thank the UOS for providing facilities and equipment for this research work. The authors would like to thank the editorial office of the journal of International Journal of Recent Contributions from Engineering, Science & IT (iJES) for processing and reviewing of the manuscript.

## 7 Authors

**Danial Abdulkareem Muhammed** received the B.Sc. degree in the field of computer science from the University of Sulaimani, Iraq, in 2009, and the master's degree (Hons.) in the field of software systems and internet technology from the University

of Sheffield, U.K., in 2012. He is currently pursuing the Ph.D. degree with the University of Sulaimani, working in the field of Artificial Intelligence in the subject of Building Evacuation under Different Situations. He worked as an Assistant Programmer for nearly two years, and then continued his study in the University of Sheffield. After that, he joined the Department of computer science, University of Sulaimani, where he is currently affiliated with.

**Soran AB. M. Saeed** received the PhD degree in Computer Science (AI-CBR) 2006 University, London, UK. In 2001, He was representative of Sulaimani University for Scientific and Cultural affairs, Current position: Vice President for Scientific Affairs and Higher Education at Sulaimani Polytechnic University (SPU). He is Head the Board of eCourt System at Sulaimani Court in Iraq Developing by AKTORS company from Estonia. His research interests are Artificial Intelligence, E-Commerce, Information Security, Business Technology, Research Methodology, and Software Engineering.

**Tarik Ahmed Rashid** received the Ph.D. degree in computer science and informatics from the College of Engineering, Mathematical and Physical Sciences, University College Dublin (UCD), in 2006, where he was a Postdoctoral Fellow of the Computer Science and Informatics School, from 2006 to 2007. He joined the University of Kurdistan Hewlêr, in 2017. His research interests include three fields. The first field is the expansion of machine learning and data mining to deal with time series applications. The second field is the development of DNA computing, optimization, swarm intelligence, and nature inspired algorithms and their applications. The third field is networking, telecommunication, and telemedicine applications.